\documentclass[%
 aip,
 cha,
 amsmath,amssymb,
 reprint,%
]{revtex4-1}

\usepackage{graphicx}
\usepackage{dcolumn}
\usepackage{bm}
\usepackage{color}

\begin{document}


\title{Intrinsic speed characteristics of a self-propelled camphor disk under repulsive perturbations}

\author{Yuki Koyano}
\email{koyano@garnet.kobe-u.ac.jp}
\affiliation{Graduate School of Human Development and Environment, Kobe University, Kobe 657-0011, Japan}

\author{Jerzy G\'{o}recki}
\email{jgorecki@ichf.edu.pl}
\affiliation{Institute of Physical Chemistry, Polish Academy of Sciences, 01-224 Warsaw, Poland}

\author{Hiroyuki Kitahata}
\email{kitahata@chiba-u.jp}
\affiliation{Department of Physics, Graduate School of Science, Chiba University, Chiba 263-8522, Japan}

\date{\today}

\begin{abstract}
Camphor is a well-studied material capable of generating self-propelled motion at a water surface, and the resulting dynamics can exhibit a wide range of behaviors. Here, we analyze a one-dimensional model describing a mobile camphor disk perturbed by a second localized camphor source. The interaction between the rotor and the perturbing disk is represented by a distance-dependent potential. The study is motivated by experiments in which a camphor rotor interacts with a fixed camphor disk placed on the water surface. Numerical simulations of the model reproduce the essential features of the experimentally observed position-dependent rotor velocity for all considered forms of the potential. For weak perturbations, we derive analytical solutions valid for arbitrary potential profiles. Both the simulations and the analytical results demonstrate a pronounced asymmetry in the rotor velocity depending on whether the rotor approaches or recedes from the perturbation.
\end{abstract}


\maketitle 

\begin{quotation}

Studies on active matter have been attracting scientific attention as a fascinating field of research that bridges physics, chemistry, biology, and mathematics, offering insights into both natural phenomena and potential technological innovations. Systems that include active objects operate in far-from-thermodynamic equilibrium conditions, requiring new ideas for their modeling because they challenge classical concepts of interactions, such as reciprocality. Self-propelled motion (autonomous movement generated by individual entities) on the liquid surface is one of the characteristic phenomena observed in active matter. It seems especially interesting because it can drive biological complexity and inspire cutting-edge technologies such as targeted drug delivery or programmable materials that self-assemble on demand. The development of mathematical models describing the spatiotemporal evolution of active matter systems is necessary, particularly for those comprising large numbers of interacting self-propelled agents.
Existing realistic models are often computationally demanding, which limits their applicability to complex systems involving multiple interacting components. Experimental systems that exhibit intrinsic behaviors across broad parameter ranges provide valuable benchmarks for evaluating and validating theoretical and computational approaches. Here, we demonstrate that the speed of a rotating camphor boat, when influenced by a stationary camphor disk fixed in position, exhibits a distinct asymmetry when expressed as a function of the distance between the rotor and the perturbation. Specifically, at the same distance from the perturbing disk, the rotor moves at different speeds depending on whether it is approaching or receding from the perturbation. This directional dependence of the rotor’s velocity is quantitatively reproduced by theoretical models that describe the interaction using a distance-dependent repulsive potential. Importantly, the asymmetry in the rotor’s motion persists even when the strength of this potential is small, indicating that it represents a robust feature of the underlying interaction mechanism. This conclusion is important because it proves that the Hamiltonian description of the system cannot describe the considered experiment, regardless of the potential strength.
\end{quotation}

\section{Introduction}

Camphor is one of the most popular substances that exhibit self-propelled motion when its pieces are placed on a clean water surface~\cite{doi:10.1021/la970196p, C5CP00541H}. This surprising phenomenon was discovered over 250 years ago ~\cite{Skey1878,doi:10.1098/rspl.1860.0124,doi:10.1098/rspl.1889.0099}, and it has been intensively studied. The experiments can be simple, inexpensive, and safe, yet the time evolution of camphor objects is complex and exhibits many interesting types of behaviors. In many cases, explaining observed effects represents a significant challenge for modeling and simulation. Thermodynamically, the system is dissipative, which means that camphor molecules released from camphor pieces form a surface layer and evaporate to the environment. The surface tension of water is a decreasing function of camphor surface concentration. The driving force results from surface tension inhomogeneities around the camphor piece. Such inhomogeneities can appear from fluctuations in the surface concentration of camphor molecules. The kinetic energy of the camphor piece is dissipated via hydrodynamic drag. Depending on the geometries of the camphor piece and the water surface, we can observe translational motion~\cite{doi:10.1021/la970196p,PhysRevE.87.010901}, rotational motion~\cite{doi:10.1021/la970196p,PhysRevE.99.022211}, or different types of complex motion~\cite{KoreaSwimers,Kitahata_PCCP2022}. A synchronized motion of multiple camphor disks is one of the interesting types of time evolution~\cite{Kohira_Langmuir,Suematsu2015,EI201810,PhysRevE.99.012204}. In such cases, the objects share the concentration field of camphor molecules at the water surface and interact through it.

The motion of camphor objects can be described by a model that includes equations of motion for objects and hydrodynamic flow coupled with the equation for the concentration of camphor molecules at the water surface. A significant simplification of numerical complexity can be achieved if the hydrodynamic flows and transport are approximated by a single reaction-diffusion equation~\cite{doi:10.1021/jp004505n,NAGAYAMA2004151,book-chap2}. The driving force in the equation of motion depends on the gradient of camphor concentration around the object. The source term in the time evolution equation for the camphor concentration is localized at the positions of camphor disks because they continuously release camphor molecules. Many examples illustrate that such an approximation gives a realistic description of phenomena observed in experiments~\cite{Nakata2016,Kitahata_PCCP2022}. In special cases, the model for the complex dynamics of camphor objects can be reduced to ordinary differential equations describing the evolution close to the solution for the rest state~\cite{PhysRevE.94.042215,PhysRevE.99.022211,book-chap2} or for a rotating state~\cite{Koyano_PhysRevE.96.012609,PhysRevE.108.024217}. The resulting equations are simple enough to be solved analytically. It has been demonstrated that the essential features of self-propelled motion are correctly reflected by solutions of reduced equations. It means that, for a solution describing self-propulsion, there is a basin of parameter values in which velocity and type of trajectory obtained by the reduced equations match the complete reaction-diffusion model of object motion.

An alternative mathematical model for the evolution of multiple camphor-propelled objects has been proposed~\cite{Soh}. The model uses binary potentials to represent interactions between different objects. For a given pair of objects, the corresponding interaction potential depends only on the distance between the objects and does not include other variables. The forces are calculated from the potential gradient~\cite{Soh, PhysRevE.106.024201, PhysRevE.108.024217, PhysRevE.110.064208}. Different functions have been applied to describe interaction potentials starting from a simple exponential decay~\cite{Soh}, Yukawa-type potentials~\cite{PhysRevE.99.012204,PhysRevE.101.052202,PhysRevE.103.012214,PhysRevE.106.024201,PhysRevE.108.024217} and ending on combinations of Coulomb and exponential functions~\cite{PhysRevE.110.064208}. Moreover, there are reports in which the evolution model is simplified, assuming that the total energy of self-propelled objects was conserved. It means that the hydrodynamic drag forces are neglected. Such assumptions allow for a significant reduction in the numerical complexity of models. In our opinion, the validity of energy conservation can be questioned, keeping in mind that camphor-water systems are energy- and mass-dissipative.

Systems showing a specific character of the time evolution are good candidates to test simplifications in models. Among them, those with inherent properties that are held in a wide range of system parameters are especially useful for model verification. Here, we focus our attention on a simple system that shows an inherent asymmetry of object speed.
We present the analysis of the model describing the time evolution of a rotor propelled by a camphor disk, which is perturbed by another identical camphor disk fixed at a specific position on the water surface. Especially, we discuss the velocity modulation dynamics when the camphor disk passes through the perturbation. The choice of such a system is inspired by our recent experiments~\cite{gorecki2025modelingtimeevolutioncamphor}, briefly described in the next section, that demonstrated a specific asymmetry of the rotor speed as a function of the distance from the perturbation. The one-dimensional mathematical model presented in Section~\ref{sec:model} used the potential to describe interactions between the rotor and the perturbing disk and included the drag term. Numerical simulations discussed in Section~\ref{sec:numerical} confirmed that such a model correctly reproduces the character of the rotor velocity as a function of its distance from the perturbation. Finally, in Section~\ref{sec:analysis}, we analyze the evolution equations considering the expansion in the strength of interaction. We present the analytical solution of evolution that includes potential-dependent parameters. Based on the expansion, we prove that the asymmetry of rotor velocity represents an inherent feature of the system because it retains its character for all small-amplitude repulsive perturbation potentials.

\section{Experimental inspiration for the investigated model \label{sec:experiment}}

\begin{figure}
    \centering
    \includegraphics{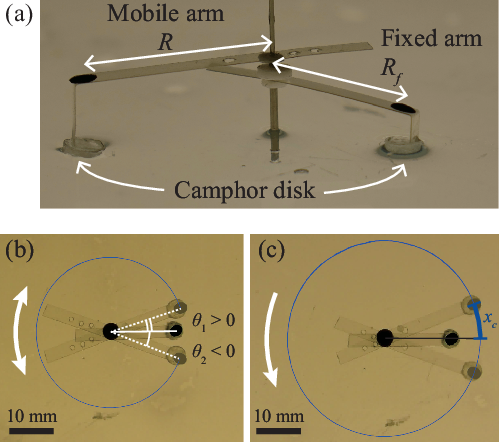}
    \caption{Experiments illustrating a mobile arm (rotor) propelled by a camphor disk and perturbed by a fixed camphor disk placed at a distance $R_f \cong 15~\mathrm{mm}$ from the axis. (a)~Geometry of the setup. (b,c)~ Two overlapped snapshots of rotor positions around the perturbation. The blue arc or circle shows the trajectory of the black marker, and arrows indicate the direction of motion. The angle $\theta=0$ corresponds to the location of the fixed camphor disk. (b)~Arm reflection for $R \cong 17~\mathrm{mm}$. The time difference between rotors below and above the fixed camphor disk is $1.14~\mathrm{s}$. The mobile camphor disk was reflected at the angles $\theta_1 \sim 0.30~\mathrm{rad}$ and $\theta_2 \sim -0.33~\mathrm{rad}$. (c) Continuous rotation for $R \cong 22~\mathrm{mm}$. The time difference between the lower and upper rotor positions is $0.4~\mathrm{s}$. The length of the thick trajectory segment indicates $x_c$ for the upper rotor position.}
    \label{fig_exp1}
\end{figure}

\begin{figure}
    \includegraphics[width=8.5cm]{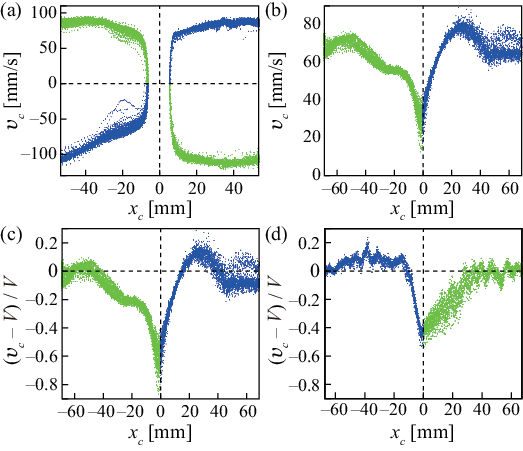}
    \caption{ 
    Rotor velocity $v_c(x_c)$ as a function of the propelling disk position $x_c$. The green and blue points mark results for a rotor moving towards and away from the perturbation, respectively. (a)~$15900$ points representing velocities measured during $318~\mathrm{s}$ (almost $130$ periods) long observation of rotor strongly perturbed by the fixed disk ($R \cong 17~\mathrm{mm}$). (b)~$10500$ points representing velocities measured during $210~\mathrm{s}$ (almost $100$ rotations) long observation of the rotor weakly perturbed by the fixed disk ($R \cong 22~\mathrm{mm}$). (c)~Scaled difference between the local velocity $v_c(x_c)$ and its stationary value $(v_c(x_c)-V)/V$ calculated form the data shown in (b). Here $V= 70~\mathrm{mm/s}$. (d) as (c) but for perturbation of rotor rotating clockwise ($R \cong 21.4~\mathrm{mm}$). The data were collected over a duration of $90~\mathrm{s}$, during which the rotation period was approximately $1.5~\mathrm{s}$ and $V= 98~\mathrm{mm/s}$.}
    \label{fig_exp2}
\end{figure}

Experiments with a camphor rotor perturbed by a camphor piece immobilized on the water surface have inspired the theoretical study presented in Secs.~III-V. The experimental system is shown in Fig.~\ref{fig_exp1}(a). Two camphor disks with a diameter of $4~\mathrm{mm}$ were located on the water surface. They were attached at the ends of coaxial plastic arms with radii of $R$ and $R_f$, respectively. The arms were lifted above the water and had no contact with the surface. For the digital processing of experimental results, the locations of both disks were marked with black dots. The shorter arm was fixed at a given position. In the experimental results reported below, $R_f$ is fixed at $15~\mathrm{mm}$. The other longer arm was allowed to rotate around the vertical axis. 
In experiments, we observed the rotation of the longer arm. Our previous studies of the self-propelled motion of camphor disks and rotors on a water surface \cite{Koyano_PhysRevE.96.012609} demonstrated stable dynamics over timescales of up to $30$~minutes. In test experiments with a single rotor of radius $R \cong 20$~mm, the rotation speed decreased by no more than 3\% during the first 10~minutes of the experiment, which was sufficient to observe the effects discussed below. If required, the duration of stable operation can be extended to more than $1$~hour using pills composed of camphene–camphor–polypropylene plastic \cite{molecules26113116}.
Figures~\ref{fig_exp1}(b,c) illustrate overlapped snapshots from two experiments together with trajectories of the rotating arms. We describe the motion in the reference frame in which the rotation axis is at the origin of the coordinate system and the fixed disk is located on the positive branch of the horizontal axis. The blue arcs in Figs.~\ref{fig_exp1}(b,c) show the trajectory of the rotating arm. The rotor position can be defined by the angle $\theta \in [-\pi, \pi]$ between the directions of the longer rotating and shorter immobile arms. In our convention, the positive angular velocity corresponds to a counterclockwise direction. Alternatively, the rotor position can be expressed as $x_c = R \theta \in [-R\pi, R\pi]$, representing the signed arc length measured along the trajectory from the horizontal axis to the rotor. The rotor speed does not exceed $100~\mathrm{mm/s}$ as presented in Fig.~\ref{fig_exp2}. The corresponding Reynolds number is below $400$, indicating that the flow remains almost in the laminar regime.

Figure~\ref{fig_exp1}(b) shows the trajectory for a strong perturbation ($R \cong 17~\mathrm{mm}$). It has an arc form with the angle $\sim 5.65~\mathrm{rad}$. The rotating arm was reflected after the interaction with the fixed disk. The rotor velocity $v_c(x_c)$ as a function of $x_c$ is presented in Fig.~\ref{fig_exp2}(a). The 15900 points represent velocities measured during $318~\mathrm{s}$ (almost $130$ periods) long observation of the rotor evolution. The green and blue points mark velocities for the rotor moving towards and away from the perturbation, respectively. We can notice that points representing $v_c(x_c)$ do not show an anti-symmetrical relation with respect to $x = 0$. For positions close to the perturbation, the speed of the rotor moving towards the perturbation is larger than that for the rotor moving away. It means that there is no function $E_p(x_c)$ that, when it is added to the position-dependent kinetic energy, returns a constant value independent on $x_c$. Therefore, the energy-conserving (Hamiltonian) model with a distance-dependent potential \cite{PhysRevE.106.024201, PhysRevE.108.024217, PhysRevE.110.064208} does not correctly describe the time evolution of a strongly perturbed system.

Figure~\ref{fig_exp1}(c) illustrates the case in which the perturbation of the rotating disk by the fixed one is moderate ($R \cong 22~\mathrm{mm}$). The rotor could make a full circle (here in counterclockwise direction) and slowed down close to the location of a fixed disk. The rotor velocity $v_c(x_c)$ as a function of rotor position $x_c$ on the trajectory is presented in Fig.~\ref{fig_exp2}(b). $10500$ points represent velocities measured during $210~\mathrm{s}$ (over $100$ rotations) long experiment. Here, $v_c(x_c)$ is not a symmetrical function of $x_c$. The speed of the rotor moving away from the perturbation has a characteristic maximum exceeding the rotor speed $V$ at a long distance from the perturbation. 
The asymmetry of velocity is clearly seen if we introduce the scaled variable $(v_c(x_c)-V)/V$, and the corresponding data are shown in Fig.~\ref{fig_exp2}(c). A qualitatively similar asymmetric speed as a function of the distance from the perturbation is observed for a weakly perturbed rotor rotating clockwise. Figure~\ref{fig_exp2}(d) shows the distance-dependent rotor velocity for $R \cong 21.4~\mathrm{mm}$. Here, the data were collected over a duration of $90~\mathrm{s}$, during which the rotation period was approximately $1.5~\mathrm{s}$.

In both cases presented above, the minimum distance between the camphor disk propelling the rotor and the camphor disk acting as the perturbation exceeded $2~\mathrm{mm}$. Thus, the observed interaction is mediated solely by the surface camphor concentration between the two disks. We also performed control experiments in which a Teflon cylinder of diameter $5~\mathrm{mm}$ was used as the perturbation. Such a cylinder acts as a passive obstacle that does not release surface-active molecules.
For the geometry shown in Fig.~\ref{fig_exp1}(c) ($R \cong 22~\mathrm{mm}$, $R_f = 15~\mathrm{mm}$), no change in rotor speed was observed as the rotor passed near the cylinder. For the geometry similar to Fig.~\ref{fig_exp1}(b) ($R \cong 19~\mathrm{mm}$, $R_f = 15~\mathrm{mm}$), mechanical collisions between the rotor and the cylinder occurred. Upon collision, the rotor speed dropped instantaneously to zero, and after a short delay the rotor began to move in the opposite direction. As expected, the resulting speed profile $v_c(x_c)$ in this case is strongly asymmetric.

\section{Mathematical model \label{sec:model}}

Now let us discuss a one-dimensional mathematical model motivated by the experimental results described in Sec.~II. Because the camphor rotor moves along a circular trajectory, we introduce a coordinate $x$ defined along this path (cf. Fig.~\ref{fig_exp1}(c)). For simplicity, the domain of system evolution is represented as a one-dimensional interval $\Omega = [-L/2, L/2]$, where the interval length is $L=2 \pi R$. 
The coordinates are chosen such that the fixed camphor disk (the perturbation) is located at $x = 0$. The position of one-dimensional mobile disk is denoted by $x_c(t)$, and $u(x,t)$ represents the concentration field generated by camphor dissipation from this disk.

The rotor motion is described by the Newtonian equation:
\begin{align}
m \frac{d^2x_c}{dt^2}= - \eta \frac{dx_c}{dt} + F - \left.\frac{dU}{dx}\right|_{x = x_c},
\label{eq-motion}
\end{align}
where $m$ and $\eta$ are the mass and friction coefficient of the camphor disk. Equation~\eqref{eq-motion} is solved with periodic boundary conditions at the ends of $\Omega$. $U(x)$ is the potential describing the force that originates from the surface tension gradient generated by camphor molecules dissipated from the fixed camphor disk acting as a perturbation. $F$ is the force arising from the camphor concentration field $u(x,t)$, which can be expressed as ~\cite{AdamsonGast1997, book-chap2}:
\begin{align}
F = \Gamma \int_\Omega \frac{\partial u}{\partial x} \Theta\left(r - \left|x - x_c\right|\right) dx
\end{align}
where $\Theta$ is the Heaviside function, i.e. $\Theta(z) = 1$ for $z \geq 0$ and $\Theta(z) = 0$ for $z < 0$. $\Gamma$ is a positive constant which relates the surface tension with the surface camphor concentration \cite{Suematsu_Langmuir}. Thus we obtain:
\begin{align}
    F = -\Gamma \left[ u(x_c + r) - u(x_c - r) \right].
    \label{eq-force}
\end{align}

In our numerical model, we do not explicitly compute the Marangoni flows forcing the camphor transport, as their full three-dimensional description for moving sources is highly complex. Instead, the time evolution of the concentration field $u(x,t)$ is approximated by a reaction–diffusion equation:
\begin{align}
    \frac{\partial u}{\partial t} = D \frac{\partial^2u}{\partial x^2} - a u + S(x - x_c(t)).
    \label{eq-u}
\end{align}
The first, second, and third terms in the right side represent camphor transport at the water surface, evaporation to the air, and the inflow of camphor molecules from the disk. Here, $D$ is the effective diffusion coefficient that includes the normal diffusion and convective transport~\cite{10.1063/1.5021502,Suematsu_Langmuir} and $a$ is the rate of evaporation and dissolution. The camphor source term is explicitly represented as:
\begin{align}
    S(x) = \frac{S_0}{2r}\Theta\left( r - \left| x\right| \right), \label{eq_S}
\end{align}
where $S_0$ is the supply rate and $2r$ is the size (diameter) of the camphor disk.

The one-dimensional model based on Eqs.~\eqref{eq-motion} and \eqref{eq-u} is a simplification of a more realistic two-dimensional model of the system considered in Ref.~\cite{gorecki2025modelingtimeevolutioncamphor}. Numerical simulations of the two-dimensional model show good agreement with the experimentally measured speed profile in the vicinity of a weak perturbation [see Fig.~5(c) of Ref.~\cite{gorecki2025modelingtimeevolutioncamphor}]. However, the two-dimensional model can be solved only numerically and requires the specification of a particular set of parameters, whereas the one-dimensional model allows for the analytical solution presented in Section~\ref{sec:analysis}.

The model can be transformed into the dimensionless form using the units of space, time, concentration, and force as defined as follows: the length unit equals the diffusion length $\sqrt{D/a}$, the time unit is given by $1/a$, the concentration unit is defined by $S_0/\sqrt{Da}$, and the unit of force by $\Gamma S_0 / \sqrt{Da}$.

With the dimensionless variables, the time evolution equations have the form:
\begin{align}
\frac{\partial u}{\partial t} = \frac{\partial^2u}{\partial x^2} - u + \frac{1}{2r} \Theta \left( r - \left|x - x_c \right| \right), \label{equ}
\end{align}
\begin{align}
    m \frac{d^2x_c}{dt^2} = -\eta \frac{dx_c}{dt} - \left[u(x_c + r) - u(x_c - r)\right] -\left.\frac{dU}{dx}\right|_{x = x_c}. \label{eqm}
\end{align}
Noting that objects floating on the water surface are repelled from regions of higher camphor concentration, we introduce a repulsive potential $U(x)$ with a single maximum located at $x=0$.
It means that $U(x)$ is a non-decreasing function in $[-L/2,0)$ and a non-increasing function in $(0,L/2]$. In the following, we describe $U(x)$ as:
\begin{align}
    U(x) = U_0~ \mathcal{U}(x).
\end{align}
where $U_0$ is the maximum of potential and the potential profile is given by a function $~\mathcal{U}(x)$ such that $\mathcal{U}(0) = 1$.

As examples of potential profiles $\mathcal{U}(x)$, we consider the following four functions: the Gaussian $\mathcal{U}_G(x)$, exponential $\mathcal{U}_e(x)$, piecewise linear $\mathcal{U}_1(x)$, and piecewise quadratic $\mathcal{U}_2(x)$. The explicit forms of them are described as follows:
\begin{align}
    \mathcal{U}_G(x) = \exp \left( - \pi x^2 \right), \label{UG}
\end{align}
\begin{align}
    \mathcal{U}_e(x) = \exp \left( - 2\left|x\right|\right), \label{UE}
\end{align}
\begin{align}
    \mathcal{U}_1(x) =  \left\{ \begin{array}{ll} 1 - \left|x\right|, & \left|x\right| < 1, \\ 0. & \mathrm{otherwise}, \end{array} \right. \label{U1}
\end{align}
\begin{align}
    \mathcal{U}_2(x) =  \left\{ \begin{array}{ll} 1 - 2 x^2, &  \left|x\right| \leq \dfrac{1}{2}, \\ 2\left(\left|x\right| - 1 \right)^2,  &  \dfrac{1}{2} < \left|x \right| < 1,\\ 0, & \left|x \right| \geq 1. \end{array} \right. \label{U2}
\end{align}
All these potentials are normalized as follows:
\begin{align}
    \int_{-\infty}^\infty \mathcal{U}(x) dx = 1.
\end{align}

\section{Numerical simulations \label{sec:numerical}}

\begin{figure}
    \centering
\includegraphics{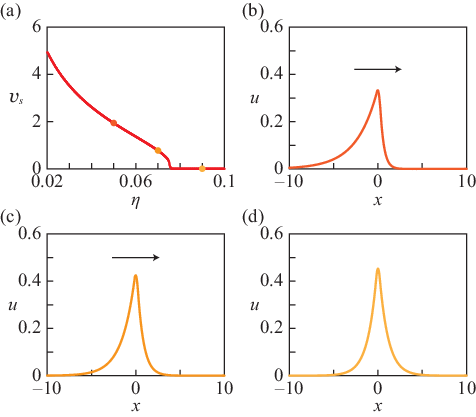}
 \caption{Numerical simulation results of the steady state for a non-perturbed case, i.e. $U(x) \equiv 0$. (a)~Stationary speed $v_s$ of the camphor disk as a function of $\eta$. (b-d)~The steady-state surface camphor concentration profiles $u(x)$ for $\eta = 0.05$(b), $0.07$(c), and $0.09$(d). In all cases, the camphor disk was located at $x_c = 0$. For $\eta = 0.09$ the disk remained stationary at this position. For $\eta = 0.05$ and $0.07$ it was moving in the positive $x$ direction with constant speeds $V = v_s(\eta = 0.07) = 0.777$ and $v_s(\eta = 0.05) = 1.94$.}
    \label{fig_sim1}
\end{figure}

We performed numerical simulations using the mathematical model introduced in Section~\ref{sec:model}. To avoid numerical instability, we introduced a smooth function $Q(x)$ describing the moving pill instead of the Heaviside function $\Theta(x)$ in Eq.~\eqref{eq_S}:
\begin{align}
    Q(x) = \frac{1}{2}\left( 1 + \tanh \frac{x}{\delta} \right).
\end{align}
 Here, $\delta$ is the smoothing parameter and $\displaystyle{\Theta(x)= \lim_{\delta \to +0}} Q(x)$. Applying the source term defined by $Q(x)$, the equation of motion (Eq.(\ref{eq-motion})) has the form:
\begin{align}
     m \frac{d^2x_c}{dt^2} = -\eta \frac{dx_c}{dt} - \int_\Omega \frac{\partial u}{\partial x} Q\left(r - \left|x - x_c\right|\right) dx - \left.\frac{dU}{dx}\right|_{x=x_c}.
\end{align}
which was used in our numerical simulations. Here, $\Omega$ denotes the whole calculation domain $\Omega = [-L/2, L/2]$.

The numerical simulations were performed by the explicit scheme. The time evolution was calculated by the Euler method with the time step of $\Delta t = 0.0001$ and the spatial mesh $\Delta x = 0.05$. The parameters were set as $m =0.01$, $L = 50$, $r = 0.2$, and $\delta = 0.03$. The initial conditions were $u(x,t=0) \equiv 0$, $x_c(t=0) = -L/2$, and $dx_c/dt|_{t=0}= 0.1$.

\begin{figure}
    \centering
\includegraphics{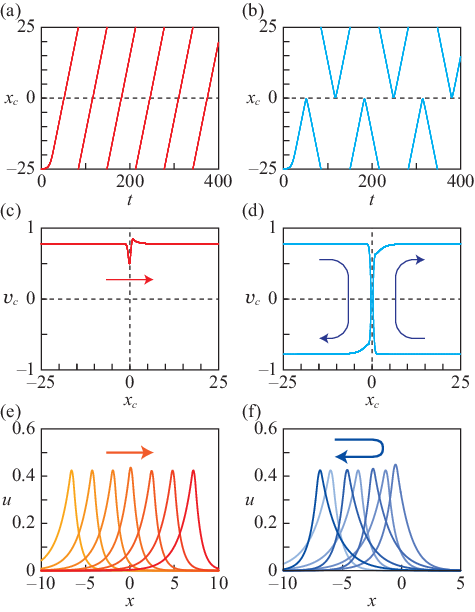}
 \caption{Numerical simulation results on the effect of the potential amplitude on the camphor disk motion. Here, we adopted the Gaussian potential $\mathcal{U}_G(x)$ (Eq.~\eqref{UG}). (a,b)~Time series of the camphor disk position $x_c$ for $U_0 = 0.005$(a) and $0.01$(b). Unidirectional motion and reciprocal motion were observed in (a) and (b), respectively. (c,d)~Camphor disk velocity $v_c = d x_c/dt$ as a function of its location $x_c$. (c) and (d) correspond to (a) and (b), respectively. (e,f)~Profiles of the concentration field $u(x)$ in the case of unidirectional and reciprocal motions corresponding to (c) and (d), respectively. The time difference between subsequent profiles is 3 time unit.}
    \label{fig_sim2}
\end{figure}

\begin{figure}
    \centering
    \includegraphics{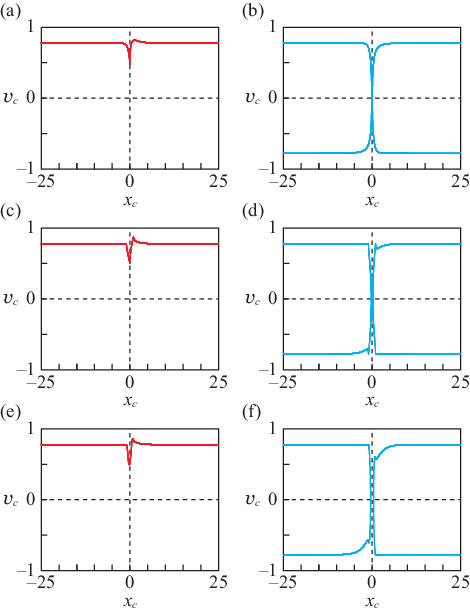}
 \caption{Numerical simulation results with different forms of the potential and their amplitude. The camphor disk velocity $v_c = d x_c/dt$ is shown as a function of its location $x_c$. (a,b)~Exponential potential $\mathcal{U}_e(x)$  (Eq.~\eqref{UE}). (c,d)~Piecewise linear potential $\mathcal{U}_1(x)$ (Eq.~\eqref{U1}). (e,f)~Piecewise quadratic potential $\mathcal{U}_2(x)$ (Eq.~\eqref{U2}). The unidirectional motions for $U_0 = 0.005$(a,c,e) and the reciprocal motions for $U_0 =0.01$(b,d,f) are shown.}
    \label{fig_sim3}
\end{figure}

\begin{figure}
    \centering
    \includegraphics{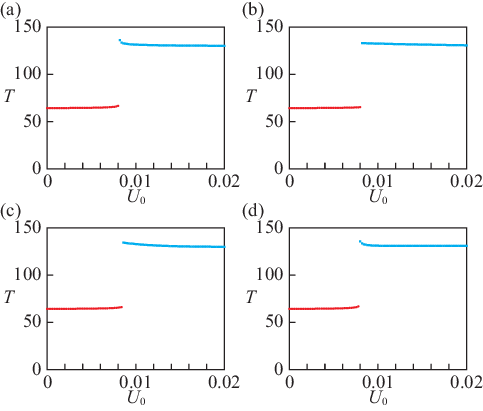}
 \caption{Numerical simulation results of the period $T$ as a function of the potential amplitude $U_0$. The period was measured by the time interval between two successive crossing of a selected trajectory point in the $x_c$-$v_c$ plane. The red and cyan lines show the unidirectional motion and reciprocal motion, respectively. (a)~Gaussian potential $\mathcal{U}_G(x)$. (b)~Exponential potential $\mathcal{U}_e(x)$. (c)~Piecewise linear potential $\mathcal{U}_1(x)$. (d)~Piecewise quadratic potential $\mathcal{U}_2(x)$.}
    \label{fig_sim4}
\end{figure}

First, we considered the time evolution of a non-perturbed system ($U(x) \equiv 0$) for various values of $\eta$. The stationary speed $v_s$ of the camphor disk as a function of $\eta$ is plotted in Fig.~\ref{fig_sim1}(a). For $\eta > 0.076$, $v_s$ was zero, whereas it had a finite value for $\eta < 0.076$, which suggests a supercritical bifurcation. Hereafter, we adopted $\eta = 0.07$ since we focused on the case close to the bifurcation point. The steady-state surface camphor concentration profiles $u(x)$ for $\eta = 0.05$(b), $0.07$(c), and $0.09$(d) are shown in Figs. \ref{fig_sim1} (b), (c) and (d), respectively. For $\eta = 0.09$, the disk remained stationary at $x_c=0$. For $\eta = 0.05$ and $0.07$, the disk was moving in the positive $x$ direction with constant speeds $V = v_s(\eta = 0.07) = 0.777$ and $v_s(\eta = 0.05) = 1.94$, respectively. As noted in Ref.~\cite{Soh}, for a stationary disk the concentration profile $u(x)$ is symmetric with respect to the disk position. In contrast, for a moving disk $u(x)$ becomes asymmetric, exhibiting a rapid decrease in camphor concentration in front of the disk and a more gradual decay behind it. This asymmetry increases with disk speed. While the maximum value of $u(x)$ decreases slightly as the disk speed increases, this effect remains small.

Next, we investigated the influence of the potential on the camphor disk motion. The amplitudes of the considered potentials were scaled by a parameter $U_0$. For the Gaussian potential in Eq.~\eqref{UG} with a small amplitude ($U_0 = 0.005$), the camphor disk moved in one direction and the speed was modulated around the potential maximum at $x = 0$ as shown in Figs.~\ref{fig_sim2}(a,c). For the larger amplitude ($U_0 = 0.01$), the disk is unable to cross the potential barrier; instead, it is reflected and undergoes reciprocal motion, as illustrated in Fig.~\ref{fig_sim2}(b). The corresponding position-dependent velocity for this motion is shown in Fig.~\ref{fig_sim2}(d). The temporal evolution of the surface camphor concentration $u(x,t)$ released by the rotor is shown in Figs.~\ref{fig_sim2}(e) and \ref{fig_sim2}(f) and provides insight into the observed asymmetry in rotor speed (cf. Fig.~\ref{fig_exp2}). In the case of weak perturbation [Fig.~\ref{fig_sim2}(e)], as the disk approaches the perturbation, the camphor concentration gradually increases in front of the disk due to camphor released from the perturbing source, leading to a reduction in speed. After passing the perturbation, the camphor concentration remains low in front of the disk and becomes elevated behind it, because camphor released from the perturbation adds to that emitted by the disk. As a result, the disk speed increases rapidly and can exceed the unperturbed steady speed $V$.

In contrast, for strong perturbations the disk speed decreases rapidly upon approach, but the concentration profile $u(x)$ must reverse in order to drive motion in the opposite direction. In this case, camphor released from the perturbation increases the concentration in the region between the disk and the perturbation. Consequently, the increase in disk speed after contact with the perturbation is significantly slower than the speed decrease observed during approach.

We also performed numerical simulations for other types of potentials and found that the unidirectional motion was exhibited for a small perturbation ($U_0 = 0.005$) while the reciprocal motion was exhibited for a large one ($U_0 = 0.01$). The trajectories at $U_0 = 0.005$, $0.01$ for the cases with the exponential, piecewise-linear, and piecewise-quadratic potentials are shown in Fig.~\ref{fig_sim3}.

Figures~\ref{fig_sim2} and \ref{fig_sim3} show that the model correctly reproduces the asymmetry of the camphor disk velocity close to the potential. For a small $U_0$, where rotation was observed, the camphor disk velocity decreased fast when it approached the potential and increased above the stationary velocity $V$ when it moved away. Moreover, in all simulations, the minimum speed was observed before the disk reached the potential center. For the larger $U_0$, when the motion was reciprocal, there was a clear difference between the fast decrease in speed when the disk approached the potential and its slower increase when the disk moved away. The same character of speed as a function of the distance from the potential center was observed in experiments as shown in Fig.~\ref{fig_exp2}.   

In order to clearly show the transition between the unidirectional motion and reciprocal motion, we changed the amplitude of the potential $U_0$ and investigated the type of motion and the period $T$. The period was measured by the time interval between two successive crossing of a selected trajectory point in the $x_c$-$v_c$ plane. The results are shown in Fig.~\ref{fig_sim4}. Considering that the camphor disk was reflected close to the center of the potential $x=0$, the period of the reciprocal motion was almost the double of the one for the unidirectional motion, suggesting that the camphor disk moves with a steady-state speed in the region far from the potential. For the selected parameter values the transition between these two motions occurred around $U_0 = 0.008$ for all four types of potentials.

\begin{figure}
    \centering
    \includegraphics{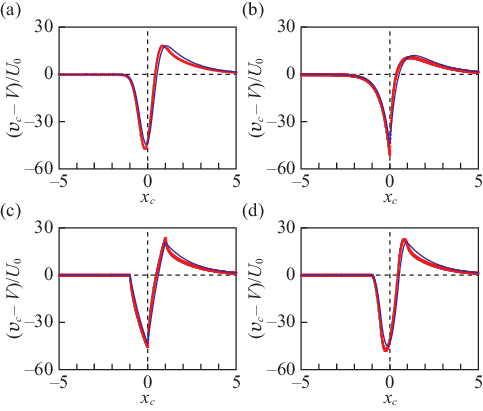}
 \caption{Comparison between numerical simulations and analytical formula for space-dependent camphor disk velocity around the perturbation for potentials of different forms. $(v_c(x_c) - V)/U_0$ is plotted as a function of $x_c$ to demonstrate that the deviation of speed is proportional to the amplitude of the potential $U_0$ for small $U_0$. The following four potential shapes were used for $\mathcal{U}(x)$. (a)~Gaussian potential $\mathcal{U}_G(x)$. (b)~Exponential potential $\mathcal{U}_e(x)$. (c)~Piecewise linear potential $\mathcal{U}_1(x)$. (d)~Piecewise quadratic potential $\mathcal{U}_2(x)$. The simulation results are plotted by a red thick line were obtained for $U_0 = 0.001$. The theoretical prediction in Eqs.~\eqref{w1_G} to \eqref{w1_2} was plotted by a blue thin line.}
    \label{fig_sim5}
\end{figure}

Now, let us focus attention on the case with the unidirectional motion. We performed numerical simulations with various $U_0$ and confirmed that the speed deviation from the steady-state speed $V$ was proportional to the potential amplitude $U_0$ if $U_0$ was sufficiently small. Numerical simulations confirmed that $(v_c - V)/U_0$ became almost constant for $U_0 \sim 0.001$, and the nonlinear effect appeared for $U_0 \gtrsim 0.002$. For $U_0 \lesssim 0.0001$, the numerical simulation results were influenced by round-up errors. The results showing  $(v_c - V)/U_0$ as a function of $x_c$ in Fig.~\ref{fig_sim5} were obtained for  $U_0 = 0.001$. Approaching the perturbation, the speed decreased and reached the minimum value before passing through $x_c = 0$. $v_c$ was less than the stationary speed $V$ at the center of the potential $x_c = 0$. Then, the speed $v_c(x_c)$ increased and reached the maximum value slightly greater than the stationary speed $V$ and then decayed to $V$. The above properties were common for the four types of potentials, though the speed shifts as well as locations of the extrema were different depending on the shape of potential profile. 

\section{Analytical results for a small perturbation \label{sec:analysis}}

In this section, we consider the perturbation around a solution that represents a camphor disk moving at a constant speed $V > 0$. This corresponds to a moderate perturbation and the experiments shown in Fig.~\ref{fig_exp1}(c). The reciprocal motion of a self-phoretic object has been recently analyzed in the previous studies~\cite{Dean_PhysRevE,PhysRevE.94.042215}. Here, we study position-dependent velocity when the perturbation is small. We assume that the amplitude $U_0$ of the potential is on the order of a small parameter $\varepsilon$, i.e.:
\begin{align}
    U_0 = \varepsilon.
\end{align}
Figure~\ref{fig_sim5} shows that the changes in the camphor disk speed are of the same order. To analyze the perturbation of the disk speed, we introduce a co-moving frame. The camphor disk position $\tilde{x}_c(t)$ in the co-moving frame, which satisfies $x_c(t) = V t + \tilde{x}_c(t)$, can be described as $\tilde{x}_c(t) = 0$. Therefore, $\dot{x}_c(t) = V + \dot{\tilde{x}}_c(t)$ and $\ddot{x}_c(t) = \ddot{\tilde{x}}_c(t)$, where $\dot{\tilde{x}}_c(t)$ and $\ddot{\tilde{x}}_c(t)$ are of the order of $\varepsilon$. The space coordinate is also considered in the co-moving frame, $x = Vt + \tilde{x}$. The evolution equation of the concentration field $\tilde{u}(\tilde{x},t)$ in the co-moving frame is given from Eq.~\eqref{equ} as:
\begin{align}
    \frac{\partial \tilde{u}}{\partial t} = V \frac{\partial \tilde{u}}{\partial \tilde{x}}  + \frac{\partial^2 \tilde{u}}{\partial \tilde{x}^2} - \tilde{u} + \frac{1}{2r}\Theta\left(r - \left|\tilde{x}\right|\right). \label{eq_comov}
\end{align}
The solution of $\tilde{u}(\tilde{x},t)$ can be expanded with respect to $\dot{\tilde{x}}_c(t)$, $\ddot{\tilde{x}}_c(t)$, and higher derivatives. We can also calculate the force expanding with respect to these quantities~\cite{PhysRevE.94.042215,Koyano_PhysRevE.96.012609,PhysRevE.99.022211}. It is explicitly described as:
\begin{align}
  F =f_0 + f_1 \dot{\tilde{x}}_c(t) + f_2 \left\{\dot{\tilde{x}}_c(t)\right\}^2 + f_{01} \ddot{\tilde{x}}_c(t) + \cdots, \label{eq_F}
\end{align}
where
\begin{align}
    f_0 =&  \frac{V}{2r \mathcal{V}}+ \frac{(1- e^{2rV})\mathcal{V}  - (1 + e^{2rV})V }{4 r \mathcal{V}}  e^{-r(V+\mathcal{V})}, \label{eq_f0}
\end{align}
\begin{align}
f_1 =&\frac{2}{r \mathcal{V}^3} -  \frac{(1 + e^{2rV})(1 + r\mathcal{V})}{r \mathcal{V}^3}e^{-r(V+\mathcal{V})}, \label{eq_f1}
\end{align}
\begin{align}
f_2 =& -\frac{3 V}{r \mathcal{V}^5}  + 
 \frac{1}{2 r \mathcal{V}^5} \left[(1+e^{2rV}) V(3+3 r \mathcal{V}+ r^2\mathcal{V}^2) \right. \nonumber \\
 &\left.+(1-e^{2rV})r\mathcal{V}^2(1+r \mathcal{V})\right ] e^{-r(V + \mathcal{V})}, \label{eq_f2}
\end{align}
\begin{align}
    f_{01} =& -\frac{6}{r\mathcal{V}^5}+ \frac{(1 + e^{2rV})(3 + 3r\mathcal{V} +r^2\mathcal{V}^2)}{r\mathcal{V}^5}e^{-r(V+\mathcal{V})}. \label{eq_f01}
\end{align}
Here, we set $\mathcal{V} = \sqrt{V^2+4}$.
The details in the derivation of these coefficients are shown in Appendix~\ref{appA}.

Using Eq.~\eqref{eq_F}, the time evolution equation for $\tilde{x}_c$ can be written as:
\begin{align}
    M \ddot{\tilde{x}}_c = f_0  - \eta \left(V + \dot{\tilde{x}}_c\right) + f_1 \dot{\tilde{x}}_c + f_2 \left\{ \dot{\tilde{x}}_c \right\}^2 - U'(Vt + \tilde{x}_c). \label{eqom}
\end{align}
where $M= m - f_{01}$ and $U'(x) = dU/dx = \varepsilon d\mathcal{U}/dx = \varepsilon \mathcal{U}'$. For the considered potentials, $\mathcal{U}'(x) \sim 0$ except a finite interval $I = [-x_I,x_I]$.

Using perturbation methods with respect to the potential amplitude $\varepsilon$, we obtain the following expression for camphor disk speed $w(x)$ as a function of the position $x$:
\begin{align}
    w(x) =& V - \varepsilon \int_{-x_I}^{x} \exp \left( - \frac{H\left(x - \xi \right)}{MV}  \right) \frac{\mathcal{U}'(\xi)}{MV} d\xi \nonumber \\ 
    & + \mathcal{O}(\varepsilon^2). \label{eq_W}
\end{align}
Here, we set $H = \eta - f_1$. The detailed derivation is shown in Appendix~\ref{appB}.

Using Eq.~\eqref{eq_W}, we can prove specific properties of $w(x)$ for a potential that is strictly repulsive and symmetric with respect to $x=0$. On the considered compact domain, we can represent such a potential as a non-negative function $\mathcal{U}(x)$. From the symmetry in the potential, we have $\mathcal{U}(x) = \mathcal{U}(-x)$ and the repulsive character of potential gives $\mathcal{U}'(x) \geq 0$ for $x \leq 0$ and $\mathcal{U}'(x) < 0$ for $x > 0$. For all such potentials, we can prove:

\noindent \textbf{Proposition 1.} The function $w(x) - V$ is neither an odd function nor an even function.\\
The derivation is given in Appendix~\ref{appC}.

\noindent \textbf{Proposition 2.} The function $w(x)$ has a minimum at $x < 0$.\\
The derivation is given in Appendix~\ref{appD}.

Finally, let us discuss the explicit form of $w(x)$ corresponding to the four examples of potentials defined by Eqs.~\eqref{UG} to \eqref{U2}.
By setting $K = H/(MV) = (\eta - f_1))/\left\{( m - f_{01})V\right\}$, the explicit forms of $w_1(x)$ for $\mathcal{U}_G(x)$, $\mathcal{U}_e(x)$, $\mathcal{U}_1(x)$, and $\mathcal{U}_2(x)$ in Eqs.~\eqref{UG} to \eqref{U2} are respectively given as:
\begin{align}
    w_1^{(G)}(x) =& \frac{K}{2MV}\exp \left( \frac{K^2}{4\pi} - Kx \right) \left[1 - \mathrm{erf}  \left( \frac{K - 2\pi x}{2\sqrt{\pi}}  \right) \right] \nonumber \\
    &- \frac{1}{MV}\exp(- \pi x^2), \label{w1_G}
\end{align}
\begin{align}
    w_1^{(e)}(x) = \left\{ \begin{array}{ll} \displaystyle{-\frac{2e^{2x}}{MV(K+2)} }, & x<0, \\ \displaystyle{\frac{2 e^{-2x}}{MV(K-2)} -\frac{4Ke^{-Kx}}{MV(K^2 - 4)}},  & x \geq 0,\end{array} \right. \label{w1_e}
\end{align}
\begin{align}
w_1^{(1)}(x) = \left\{
\begin{array}{ll}
  0, &  x \leq -1,\\
 \displaystyle{  \frac{e^{-K(x+1)}-1}{MVK}},  &  -1 < x < 0, \\
 \displaystyle{\frac{1-(2 - e^{-K}) e^{-Kx}}{MVK}}, &  0 \leq x < 1, \\
 \displaystyle{\frac{4}{MVK} \sinh^2 \left(\frac{K}{2}\right) e^{-Kx}}, & x \geq 1 . \\
 \end{array} \right. \label{w1_1}
\end{align}
\begin{widetext}
\begin{align}
w_1^{(2)}(x) = \left\{
\begin{array}{ll}
  0, &  x \leq -1,\\
  \displaystyle{-\frac{4[K(x +1)-1 + e^{-K(x+1)}]}{MV K^2}}, &  -1 < x < -\dfrac{1}{2}, \\
 \displaystyle{\frac{4[Kx - 1  + (2 e^{K/2} - 1)e^{-K(x+1)}]}{MVK^2} },  &  -\dfrac{1}{2} \leq x \leq \dfrac{1}{2}, \\
\displaystyle{\frac{4[1- K(x-1) -\left( e^{-K} + 4\sinh (K/2)
 \right) e^{-Kx}]}{MVK^2}}, &  \dfrac{1}{2} < x < 1, \\
 \displaystyle{\frac{8 \left[\sinh(K) - 2\sinh \left(K/2 \right)\right] e^{-Kx}}{MVK^2}  }, &  x \geq 1. \\
 \end{array} \right.  \label{w1_2}
\end{align}
\end{widetext}
Here, $\mathrm{erf}(z)$ is defined as
\begin{align}
    \mathrm{erf}(z) = \frac{2}{\sqrt{\pi}}\int_0^z e^{-y^2}dy.
\end{align}
It should be noted that at $K = 2$ the function $w_1^{(e)}(x)$ is not well defined by Eq.~\eqref{w1_e}. Taking the limit $K \to 2$ in this equation, we obtain:
\begin{align}
    w_1^{(e)}(x) = \left\{ \begin{array}{ll} \displaystyle{-\frac{e^{2x}}{2MV} }, & x<0, \\ \displaystyle{\frac{(4x - 1) e^{-2x}}{2MV} }, &  x \geq 0. \end{array} \right. \label{w1_ek2}
\end{align}

From the numerical results, we can show that Eq.~\eqref{Order0} almost holds; $f_0$ is numerically calculated as $f_0 = 0.548$ and $\eta V$ from the numerical simulation is $\eta V = 0.544$.
The values of $w_1$ calculated using Eqs.~\eqref{w1_G} to \eqref{w1_2}, are plotted in Fig.~\ref{fig_sim5} with blue lines. The analytical results are in a good agreement with numerical simulations.  They are slightly shifted for positive $x$, which seems to be due to the neglect of the higher-order terms.

\section{Discussion and Summary \label{sec:discussion}}

The complexity of complete models, including hydrodynamics, camphor-molecule transport, and object motion, has motivated many approximate descriptions that are supposed to reflect correctly the most important features of the observed phenomena~\cite{Lauga_Davis_2012,PhysRevFluids.5.084004,PhysRevFluids.6.104006,Ender2021,Ender2021-2}. With one of generally accepted approximations, the camphor surface concentration is represented by effective diffusion coefficient with a reaction-diffusion equation~\cite{10.1063/1.5021502,Suematsu_Langmuir} and used to calculate forces (cf. Eq.~\eqref{eq-force}). Another commonly used approach neglects the camphor concentration entirely and instead computes the interactions between self-propelled objects using a fixed, distance-dependent binary potential \cite{Soh}.

In our view, systems that exhibit characteristic time evolution over a broad range of parameters are well suited for model validation. In the present work, we focus on a system in which a mobile camphor disk (rotor) is perturbed by a camphor disk fixed at a prescribed position. For this repulsive interaction, we observe a pronounced asymmetry in the rotor speed as a function of the distance from the perturbation.

As discussed in Section~\ref{sec:experiment}, two distinct types of motion arise depending on the strength of the perturbation in the considered system. For sufficiently strong perturbations, the mobile disk is reflected after approaching the fixed one. In this regime, the rotor speed decreases rapidly as it approaches the perturbation and increases more slowly as it moves away. As shown in Section~\ref{sec:numerical}, the asymmetry observed for strong interactions is in qualitative agreement with a model that combines a reaction–diffusion equation for camphor transport, a potential representing the perturbation, and the hydrodynamic drag acting on the mobile disk.

A different type of asymmetry in the position-dependent rotor velocity near the perturbation is observed in the weak-perturbation regime [see Fig.~\ref{fig_exp2}(b,c)]. In this case, the rotor reaches its maximum speed after passing the perturbing disk. The qualitative form of the mobile disk speed as a function of the distance from the perturbation is also well captured by a model that employs a reaction–diffusion equation for camphor transport, represents the perturbation via an interaction potential, and includes the hydrodynamic drag acting on the rotor. Notably, the asymmetry persists in simulations employing different functional forms of the interaction potential. We have derived analytical results for the weak-perturbation regime that apply to arbitrary potential shapes. Numerical simulations demonstrate that, for weak potentials, the scaled velocity deviation $(v_c(x) - V)/U_0$  depends on the potential profile $~\mathcal{U}(x)$,  but is independent of its amplitude $U_0$. A quantitative comparison between the analytical expressions and numerical simulations shows excellent agreement, as illustrated in Fig.~\ref{fig_sim5}.

In deriving the analytical solution presented in Section~\ref{sec:analysis}, we considered perturbations about a steady state corresponding to a camphor disk moving at a finite velocity. In contrast, many previous studies \cite{PhysRevE.94.042215,Koyano_PhysRevE.96.012609,PhysRevE.99.022211} have focused on perturbations around a rest state, in which the camphor disk is stationary and the camphor concentration field is time independent. An expansion about the rest state can also be carried out for a weakly perturbed rotor. However, this approach generally yields lower accuracy, because the stable speed of a moving camphor disk emerges from a balance between the first- and third-order terms in the velocity obtained via perturbation theory. Consequently, the error increases as the system parameters move away from the bifurcation point.

In the present analysis, the relaxation timescale to the steady velocity, $M/H = (m - f_{01}) / (\eta - f_1)$, plays a central role. An accurate estimate of the velocity perturbation induced by the potential is obtained through direct calculations in the co-moving frame. Therefore, perturbations about the constant-velocity state provide more accurate results for describing the influence of the interaction potential. Alternatively, perturbation expansions about the rest state \cite{PhysRevE.94.042215} or Fourier-series approaches \cite{Dean_PhysRevE} may be employed to obtain analytical descriptions of reciprocal motion involving reflections at the perturbation. 

We expect that the asymmetry between the speeds of approaching and departing objects is an inherent feature of the temporal evolution of systems composed of interacting self-propelled swimmers. At low swimmer densities, the dynamics can be reasonably approximated by binary interactions, and our results provide evidence that such models naturally give rise to speed asymmetry. At higher densities, however, many-body interactions are expected to play a dominant role, and further investigations will be required to elucidate the resulting collective dynamics.

Finally, several recent studies have proposed Hamiltonian models based on energy conservation to describe the dynamics of self-propelled ribbons \cite{PhysRevE.106.024201, PhysRevE.108.024217, PhysRevE.110.064208}. These models rely on the assumption that the total mechanical energy
$E_0$ of the system is conserved. In this framework, $E_0$ is defined as the sum of the rotor kinetic energy, $I\omega(t)^2/2 = I v(t)^2/(2R^2)$, and the distance-dependent interaction potential with the perturbation, $U(r(t))$:
$
E_0 = I v(t)^2 /(2 R^2) +  U(r(t))
$.
Here, $I$ denotes the moment of inertia of the rotor, $\omega(t)$ its angular velocity, and $v(t)$ the tangential velocity.
Under this assumption, the rotor speed $|v(t)|$ depends solely on the distance from the perturbation, since
$\left|v(t)\right| = R \sqrt{2 (E_0 -  U(r(t)))/I}
$.
Consequently, $\left|v(t)\right|$ is predicted to be a symmetric function of the rotor position $x_c$. Furthermore, for a repulsive interaction potential, the model implies that the minimum speed occurs at the point of closest approach, $x_c = 0$.
As shown in Fig.~\ref{fig_exp2} and in our theoretical analysis, neither of these predictions is supported by the experimental observations or by the model that explicitly includes camphor transport and hydrodynamic drag. We therefore conclude that the Hamiltonian approach proposed in Refs.~\cite{PhysRevE.106.024201, PhysRevE.108.024217, PhysRevE.110.064208} is not applicable to the rotors investigated here, which are weakly perturbed by a stationary camphor source.

\begin{acknowledgments}
This work was supported by PAN-JSPS program ``Complexity and order in systems of deformable self-propelled objects'' (No.~JPJSBP120234601). This work was also supported by JSPS KAKENHI Grants Nos.~JP23K20815, JP24K06978, JP24K22311, JP24K16981, JP25K00918, and also the Cooperative Research Program of ``Network Joint Research Center for Materials and Devices'' (Nos.~20254003, 20251011).
This work was also supported by MEXT Promotion of Distinctive Joint Usage/Research Center Support Program Grant Number JPMXP0724020292.
\end{acknowledgments}

\section*{Data Availability Statement}

The data that support the findings of this study are available from the corresponding author upon reasonable request.

\appendix

\section{Derivation of Equations~\eqref{eq_f0}, \eqref{eq_f1}, and \eqref{eq_f01} \label{appA}}

The Green's function $G(x, t)$ for Eq.~\eqref{eq_comov} is defined as the function that satisfies:
\begin{align}
\frac{\partial G}{\partial t} = V\frac{\partial G}{\partial x} +  \frac{\partial^2 G}{\partial x^2} - G + S(x) \delta(t). \label{defGreenFunc}
\end{align}
Using the Green's function, the concentration field is obtained as:
\begin{align}
u(x, t) = \int_{-\infty}^t G(x - x_c(t'), t - t') dt'. \label{u_rt}
\end{align}
We introduce the Fourier transform of $G(x,t)$ as:
\begin{align}
    \mathcal{G}(k,t) = \int_{-\infty}^{\infty} G(x,t) e^{-ikx} dx
\end{align}
and the inverse Fourier can be calculated as:
\begin{align}
G(x,t) = \frac{1}{2\pi} \int_{-\infty}^{\infty} \mathcal{G}(k, t) e^{i kx} dk. \label{Grt}
\end{align}
From Eq.~\eqref{defGreenFunc}, we obtain:
\begin{align}
\frac{d}{dt} \mathcal{G}(k,t) = - \left(-i k V + k^2 + 1 \right) \mathcal{G}(k,t) + \mathcal{S} (k) \delta(t),
\end{align}
where
\begin{align}
\mathcal{S}(k) = \int_{\infty}^{\infty} S(x) e^{-i kx} dx.
\end{align}
Therefore, $\mathcal{G}(k,t)$ has the form:
\begin{align}
\mathcal{G}( k, t) = \left\{ \begin{array}{ll} \mathcal{S}(k) \exp \left( -K t\right), & t \geq 0, \\ 0, &t < 0, \end{array} \right. \label{Gkt-}
\end{align}
where:
\begin{align}
K = - i V k + k^2 + 1.
\end{align}

We consider the Fourier transform of $u(x, t)$ as:
\begin{align}
 \mathcal{F}(k,t) =& \int_{-\infty}^{\infty} u(x,t) e^{-ikx }dx \nonumber \\
  =& \int_{-\infty}^\infty \int_{-\infty}^t G(x- x_c(t'), t- t') dt' e^{-ikx }dx \nonumber \\
  =& \int_{-\infty}^t \int_{-\infty}^\infty  G(x- x_c(t'), t- t') e^{-ik(x - x_c(t'))}dx \nonumber \\
  & \times e^{-ikx_c(t')} dt' \nonumber \\
  =& \int_{-\infty}^t \mathcal{G}(k,t - t') e^{-ikx_c(t')} dt' \nonumber \\
=& \int_{-\infty}^t \mathcal{S}(k) e^{- K ( t - t')} e^{- i kx_c(t')} dt' \nonumber \\
=& e^{- K t} \int_{-\infty}^t \mathcal{S}(k) e^{K t'} e^{- i kx_c(t')} dt'.
\end{align}
and expand $\mathcal{F}(k, t) e^{Kt}$ using the partial integration as follows:
\begin{widetext}
\begin{align}
\mathcal{F}(k, t) e^{K t}
 & = \left[ \mathcal{S}(k) \frac{ e^{K t'}}{K} e^{ - i kx_c(t') } \right]_{-\infty}^t - \int_{-\infty}^t \mathcal{S}(k) \frac{e^{K t'}}{K} \left( - i k \dot{x}_c (t') \right) e^{- i kx_c(t')} dt' \nonumber \\
& = \left[ \mathcal{S}(k) \frac{e^{K t'}}{K} e^{ - i kx_c(t') } \right]_{-\infty}^t  -  \left[ \mathcal{S}(k) \frac{e^{K t'}}{ K^2 } \left( - i k\dot{x}_c (t') \right) e^{ - i kx_c(t') } \right]_{-\infty}^t \nonumber \\
& \qquad + \int_{-\infty}^t \mathcal{S}(k) \frac{e^{K t'}}{K^2} \left\{ - i k \ddot{x}_c (t') + \left(- i k \dot{x}_c(t') \right)^2 \right\} e^{ - i kx_c(t') } dt' \nonumber \\
& = \left[ \mathcal{S}(k) \frac{e^{K t'}}{K} e^{ - i kx_c(t') } \right]_{-\infty}^t + \left[ i k \dot{x}_c (t')  \mathcal{S}(k) \frac{e^{K t'}}{ K^2 }  e^{ - i kx_c(t') } \right]_{-\infty}^t \nonumber \\
& \qquad + \left[ \mathcal{S}(k) \frac{e^{K t'}}{K^3} \left\{ - ik \ddot{x}_c (t') + \left(- ik \dot{x}_c(t') \right)^2 \right\} e^{ - i k x_c(t') } \right]_{-\infty}^t \nonumber \\
& \qquad - \int_{-\infty}^t \mathcal{S}(k) \frac{e^{K t'}}{K^3} \left\{ - ik \dddot{x}_c (t') + 3\left(- i k \dot{x}_c(t') \right) \left(- i k \ddot{x}_c(t') \right) + \left(- i k \dot{x}_c(t') \right)^3 \right\} e^{ - i kx_c(t') } dt' \nonumber \\
& = \cdots .
\end{align}
As a result, we obtain: 
\begin{align}
\mathcal{F}(k, t) 
=&  \frac{\mathcal{S}(k)}{K} e^{- i kx_c(t)} + i \frac{\mathcal{S}(k)}{K^2} k \dot{x}_c (t)  e^{- i kx_c(t)}  + \frac{\mathcal{S}(k)}{K^3 } \left\{ - i k \ddot{x}_c (t) - \left(k \dot{x}_c(t) \right)^2 \right\} e^{ - i kx_c(t) } \nonumber \\
& + \frac{\mathcal{S}(k)}{K^4 } \left\{ i k \dddot{x}_c (t) + 3 \left(k \dot{x}_c(t) \right) \left(k \ddot{x}_c(t) \right) -i  \left( k \dot{x}_c(t) \right)^3 \right\} e^{- i kx_c(t)} + \cdots \nonumber \\
\equiv & F(k, \dot{x}_c(t), \ddot{x}_c(t), \dddot{x}_c(t)) e^{-i kx_c(t)}.
\end{align}
In real space, we get:
\begin{align}
u(x,t) =& \frac{1}{2\pi} \int_{-\infty}^{\infty} F(k, \dot{x}_c(t), \ddot{x}_c(t), \dddot{x}_c(t)) e^{i k \left(x - x_c(t) \right)} d k \nonumber \\
=& \frac{1}{2\pi} \int_{-\infty}^{\infty} \frac{\mathcal{S}(k)}{K} e^{ i k(x - x_c(t))}dk + \frac{i}{2\pi}  \dot{x}_c(t) \int_{-\infty}^{\infty} \frac{k\mathcal{S}(k)}{K^2} e^{ i k(x - x_c(t))}dk  \nonumber \\
& - \frac{i}{2\pi}\ddot{x}_c(t)  \int_{-\infty}^{\infty} \frac{k\mathcal{S}(k)}{K^3 } e^{ik(x-x_c(t))} dk - \frac{1}{2\pi} \left\{\dot{x}_c(t)\right\}^2 \int_{-\infty}^{\infty} \frac{k^2\mathcal{S}(k)}{K^3 } e^{ik(x-x_c(t))} dk
 + \cdots .\label{eq_expansion}
\end{align}
\end{widetext}
This expansion is valid for small-velocity, acceleration, and higher-order time derivatives of $x_c$.

In order to obtain the explicit expression, we apply the following procedure.
Let us consider the time evolution equation for concentration:
\begin{align}
\frac{\partial c}{\partial t} = V \frac{\partial c}{\partial x} + \frac{\partial^2 c}{\partial x^2} - \beta c + S(x - X). \label{diff_a}
\end{align}
and its stationary solution $c_0(x, X, \beta)$ that satisfies the equation:
\begin{align}
V \frac{\partial c_0}{\partial x} + \frac{\partial^2 c_0}{\partial x^2} - \beta c_0 + S(x - X) = 0. \label{stationary_eq}
\end{align}
Here, $X$ corresponds to the camphor disk position.
If the stationary solution $c_0(x, X, \beta)$ is obtained, the following equality holds:
\begin{align}
c_0(x,X, \beta) = \frac{1}{2\pi}\int_{-\infty}^{\infty} \frac{\mathcal{S}(k)}{-i V k + k^2 + \beta} e^{ik \left(x - X \right)} d k. \label{equality_trans}
\end{align}
We assume that $c_0(x, X, \beta)$ satisfies the boundary condition for $x \to \pm\infty$ and it is sufficiently smooth. The reason why we introduce new parameters $\beta$ and $X$ is that the other equalities can be obtained by considering the derivative by $\beta$ and $X$ as follows:
\begin{align}
 &\frac{\partial^{m+n} c_0}{\partial \beta^m \partial X^n} \nonumber \\
 &= \frac{1}{2\pi} \int_{-\infty}^{\infty} (-ik)^n (-1)^m m! \frac{\mathcal{S}(k)}{\left(\beta  - i Vk+ k^2 \right)^{m+1}}e^{i k \left(x - X\right)} dk.
\end{align}
By substituting $\beta = 1$ and $X= x_c(t)$, we obtain:
\begin{align}
& \left.\frac{\partial^{m+n} c_0}{\partial \beta^m \partial X^n}\right|_{X = x_c(t), \beta = 1} \nonumber \\
&=  \frac{ (-i)^n (-1)^m m!}{2\pi} \int_{-\infty}^{\infty} \frac{\mathcal{S}(k) k^n}{\left(1  - i Vk+ k^2 \right)^{m+1}} e^{ik(x-x_c)}dk,
\end{align}
and thus
\begin{align}
   &\frac{1}{2\pi}\int_{-\infty}^{\infty} \frac{k^n S(k)}{K^{m+1}} e^{ik(x - x_c(t))}dk \nonumber \\
   &=\frac{(-1)^m i^n}{m!} \left.\frac{\partial^{m+n} c_0}{\partial \beta^m \partial X^n}\right|_{X = x_c(t), \beta = 1}.
\end{align}
Using these formulae, we can calculate $u(x,t)$ as:
\begin{align}
    u(x,t) =& c_0(x,x_c(t),1) + \dot{x}_c(t) \left.\frac{\partial^2 c_0}{\partial \beta \partial X} \right|_{X= x_c(t), \beta=1} \nonumber \\
    &+ \frac{1}{2} \ddot{x}_c(t) \left.\frac{\partial^3 c_0}{\partial X\partial \beta^2}\right|_{X= x_c(t), \beta=1} \nonumber \\ & + \frac{1}{2} \left\{\dot{x}_c(t) \right\}^2 \left.\frac{\partial^4 c_0}{\partial X^2\partial \beta^2}\right|_{X= x_c(t), \beta=1} + \cdots . \label{expansion_u}
\end{align}
From the above-mentioned equations, we can calculate the concentration field in the form of an expansion with $\dot{x}_c$ and $\ddot{x}_c$. Then, the force acting on the camphor disk is calculated, and the reduced ordinary differential equation for the camphor disk motion is obtained.

Let us consider the case with a camphor disk of a finite size $2r$ in a one-dimensional finite space. If the center of the camphor disk remains at $X$ in a co-moving frame, the stationary concentration field is described as:
\begin{widetext}
\begin{align}
    c_0(x,X, \beta) = \left\{ \begin{array}{ll}
    \dfrac{\kappa_+  \sinh \left( \kappa_- r\right) }{\beta r(\kappa_+ + \kappa_-)}e^{\kappa_- (x - X)}, & x - X < - r, \\
     \dfrac{1}{2 \beta r}\left( 1 - \dfrac{\kappa_- e^{- \kappa_+ (x - X + r)} + \kappa_+ e^{ \kappa_- (x - X - r)}  }{\kappa_+ + \kappa_-}\right), & -r \leq x - X \leq r, \\
    \dfrac{\kappa_-  \sinh \left( \kappa_+  r\right) }{\beta r(\kappa_+ + \kappa_-)} e^{- \kappa_+ (x - X)}, & x - X > r, 
    \end{array}
    \right. \label{c}
\end{align}
\end{widetext}
where we define $\kappa_+$ and $\kappa_-$ as
\begin{align}
    \kappa_\pm =\sqrt{\beta + \frac{V^2}{4}} \pm \frac{V}{2}.
\end{align}
It should be noted that $\kappa_\pm > 0$.

The force working on the camphor disk is given as
\begin{align}
F=& -\left[ u(x_c + r) - u(x_c - r) \right].
\end{align}
Using Eqs.~\eqref{expansion_u} and \eqref{c}, we can calculate the force working on the camphor disk $F$ in the form of Eq.~\eqref{eq_F} with Eqs.~\eqref{eq_f0} to \eqref{eq_f01}.
It is notable that for the limit of a small source $(r \to +0)$, we get: 
\begin{align}
\lim_{r \to +0}\frac{f_0}{2r} =  \frac{V}{2 \mathcal{V}},
\end{align}
\begin{align}
\lim_{r \to +0}\frac{f_1}{2r} =  \frac{2}{\mathcal{V}^3},
\end{align}
\begin{align}
\lim_{r \to +0}\frac{f_2}{2r} =  -\frac{3V}{\mathcal{V}^5},
\end{align}
\begin{align}
\lim_{r \to +0}\frac{f_{01}}{2r} =  \frac{V^2 - 2}{2 \mathcal{V}^5}.
\end{align}

\section{Derivation of Equation~\eqref{eq_W} \label{appB}}
If $U \equiv 0$ then $\dot{\tilde{x}}_c = \ddot{\tilde{x}}_c = 0$ and Eq. \eqref{eqom} reduces to:
\begin{align}
    f_0 - \eta V = 0 \label{Order0}
\end{align}
If $\varepsilon =0$ then speed  in the co-moving frame $\tilde{v}_c(t) = \dot{\tilde{x}}_c(t)$ should equal to $0$. We assume that the dependence of $\tilde{v}_c(t)$ on  $\varepsilon$  can be approximated as:
\begin{align}
    \tilde{v}_c(t) = \varepsilon v_1(t) + \mathcal{O}(\varepsilon^2). \label{eq_tv}
\end{align}
Accordingly, $x_c(t)$ can be described as
\begin{align}
    x_c(t) = V t + \varepsilon \int_0^t v_1 (\tau) d\tau + \mathcal{O}(\varepsilon^2), \label{eq_xc}
\end{align}
where we assumed $x_c(0)=0$.

Substituting Eqs.~\eqref{eq_tv} and \eqref{eq_xc} to Eq.~\eqref{eqom}, we obtain
\begin{align}
   M\frac{d}{dt}\left(\varepsilon v_1  + \mathcal{O}(\varepsilon^2)\right) =& f_0 - \eta V - \eta \varepsilon v_1 \nonumber \\
   & + f_1 \varepsilon v_1 - \varepsilon \mathcal{U}'(Vt) + \mathcal{O}(\varepsilon^2).
\end{align}
Grouping together the terms of $\mathcal{O}(\varepsilon^1)$, we obtain: 
\begin{align}
   M \frac{d v_1}{dt} = - H v_1 - \mathcal{U}'(V t ). \label{eq1}
\end{align}
Hereafter, we assume $M=m - f_{01} > 0$ and $H = \eta - f_1 > 0$ since the motion with a constant speed $V$ becomes unstable otherwise. Then, the solution of \eqref{eq1} for the initial condition $v_1(0) = 0$ is given as:
\begin{align}
    v_1(t) = -\int_{0}^t \exp \left( - \frac{H}{M} (t - \tau) \right) \frac{\mathcal{U}'(V\tau )}{M} d\tau. \label{eq_v1}
\end{align}

To obtain the velocity profile $w(x)$ as a function of the position $x$, we assume expansion:
\begin{align}
    w(x) = V + \varepsilon w_1(x) + \mathcal{O}(\varepsilon^2). \label{eq_w}
\end{align}
The relation between $w(x)$ and $\dot{x}_c(t)$ is described as
\begin{align}
    w(x_c(t)) = \dot{x}_c(t) = V + \varepsilon v_1(t) + \mathcal{O}(\varepsilon^2).
\end{align}
By comparing the terms on the first order of $\varepsilon$ we get:
\begin{align}
    v_1(t) = w_1(x_c(t)).
\end{align}
Here, the position as a function of the time $t$ is given as:
\begin{align}
    x_c(t) =& V t + \varepsilon \int_0^t v_1 (\tau) d\tau + \mathcal{O}(\varepsilon^2), 
\end{align}
and thus we obtain:
\begin{align}
w_1(Vt) = v_1(t) + \mathcal{O}(\varepsilon).
\label{eq-32}
\end{align}
Equation~\eqref{eq-32} can be transformed to:
\begin{align}
    w_1(x) = v_1\left(\frac{x}{V}\right)+ \mathcal{O}(\varepsilon).
\end{align}
and finally, we obtain
\begin{align}
    w(x) =& V - \varepsilon \int_{-x_I}^{x} \exp \left( - \frac{H\left(x - \xi \right)}{MV}  \right) \frac{\mathcal{U}'(\xi)}{MV} d\xi \nonumber \\ 
    & + \mathcal{O}(\varepsilon^2).
\end{align}

\section{Derivation of proposition 1\label{appC}}

The velocity at $x=0$ is given as
\begin{align}
    w(0) = V - \varepsilon \int_{-x_I}^{0} \exp \left( \frac{H\xi}{MV} \right) \frac{\mathcal{U}'(\xi)}{MV} d\xi.
\end{align}
The second term is negative for a non-decreasing potential for $x \in [-x_I, 0)$ ( $\mathcal{U}'(x) \geq 0$). Therefore, $w(0) - V < 0$, which means that $w(x) - V$ is not an odd function.

From Eq.~\eqref{eq_W}, the spatial derivative of $w(x)$ equals to:
\begin{align}
    \frac{dw}{dx} =& \varepsilon \left[ \frac{H}{MV} \int_{-x_I}^x \exp \left( - \frac{H \left(x - \xi \right)}{MV} \right) \frac{\mathcal{U}'(\xi)}{MV} d\xi  - \frac{\mathcal{U}'(x)}{MV}\right] \nonumber \\
    &+ \mathcal{O}(\varepsilon^2). 
    \label{eq-36}
\end{align}
Having in mind that $\mathcal{U}'(0) = 0$,
\begin{align}
    \left. \frac{dw}{dx} \right|_{x = 0} = \varepsilon \frac{H}{MV} \int_{-x_I}^0 \exp \left(\frac{H \xi}{MV}  \right) \frac{\mathcal{U}'(\xi)}{MV} d\xi > 0 
    \label{eq-37}
\end{align}
This means that $w(x)$ and so $w(x) - V$ are not even functions. Therefore, Proposition 1 is proved. 

\section{Derivation of Proposition 2\label{appD}}

Let us notice that:
\begin{align}
    \left| H\int_{-x_I}^x \exp \left( - \frac{H \left(x - \xi \right)}{MV} \right) \frac{\mathcal{U}'(\xi)}{MV} d\xi  \right| \leq \max_{\xi \in [-x_I,x] } \left|\mathcal{U}'(\xi) \right|,
    \label{eq-36+}
\end{align}
$\mathcal{U}'(x)$ is non-negative in a compact interval $[-x_I, 0)$. Let $x_m \in [-x_I, 0)$ is the value at which $\mathcal{U}'(x)$ has a maximum. Using Eqs.~\eqref{eq-36+} and \eqref{eq-36}, we conclude that $(dw/dx)|_{x = x_m} < 0$.  On the other hand, Eq.~\eqref{eq-37} says that $(dw/dx)|_{x = 0}>0$. Therefore, there should be a minimum of $w(x)$ in $[x_m, 0]$. Its location can be found numerically by assuming that $dw/dx = 0$ in Eq.~\eqref{eq-36}.

\end{document}